\documentclass{TTP_DSL2006}
\usepackage[english]{babel}
\usepackage{array}
\usepackage{mathtext}
\usepackage{textcomp}
\usepackage{cite}
\usepackage{graphicx}
\usepackage{sidecap}
\sidecaptionvpos{figure}{c}

\newcommand{\be}{\begin{equation}}
\newcommand{\ee}{\end{equation}}

\usepackage{graphicx}
\usepackage[intlimits]{amsmath}
\usepackage{amssymb}
\usepackage{exscale}
\usepackage{hyperref}
\usepackage{times}

\renewcommand{\large}{\fontsize{14}{18pt}\selectfont}
\renewcommand{\small}{\fontsize{11}{13.6pt}\selectfont}

\newcommand{\titleformat}{\sffamily\bfseries \large}				
\newcommand{\authorformat}{\sffamily \large}						
\newcommand{\keywordsformat}{\noindent \small \sffamily}			
\newcommand{\abstractformat}{\noindent \textbf}						
\newcommand{\contentformat}{\rmfamily \normalsize\vspace{18pt}}		
\newcommand{\email}{\sffamily \small \vspace{-8pt}}					
\renewcommand{\subsection}{\textbf}									

\hypersetup{
    colorlinks,%
    citecolor=black,%
    filecolor=black,%
    linkcolor=black,%
    urlcolor=black
}

\begin{document}

\title{\titleformat Effect of mechanical stresses on the coercive force of the heterophase non-interacting nanoparticles}

\author{\authorformat Leonid Afremov \inst{1} and Yury Kirienko \inst{2}$^{,*}$}

\institute{\sffamily Far-Eastern Federal University, Vladivostok, Russia}
\maketitle

\begin{center}
\email{$^{1\,}$afremovl@mail.dvgu.ru, $^{2\,}$yury.kirienko@gmail.com, $^{*}$corresponding author}
\end{center}

\vspace{2mm} \hspace{-7.7mm} \normalsize 
\keywordsformat{\textbf{Keywords:} heterophase particles, mechanical stress, maghemite,
cobalt ferrite, elongated nanoparticles, coatings, interfacial exchange interaction.}

\contentformat

\abstractformat{Abstract.} The theoretical analysis of the effect of uniaxial stress on the magnetization of the system 
of noninteracting nanoparticles is done by an example of heterophase particles of maghemite, epitaxially coated with cobalt ferrite.
It is shown that stretching leads to a decrease in the coercive force $H_c$, and compression leads to its growth.
The residual saturation magnetization $I_{rs}$ of nanoparticles does not change. 
With increasing of interfacial exchange interaction, coercive force varies nonmonotonically.

\section{Introduction}
It is known that the reduction of size of the particle leads to an intensification of reactivity of the magnetic material. 
As a result, it is natural to assume that small particles are rather heterophase than homogeneous.
Moreover, the formation of neighboring magnetic phases can be caused by processes of oxidation 
or disintegration of the solid solution (see, eg,~\cite{Stacey1974,Gapeev1992}) occurring in the magnetically ordered grain.
Most of the ultrafine magnetic materials of practical interest is two-phase or multiphase single-domain particles. 
For example, they are carriers of information in the magnetic memory elements, also they are widely used in modern biophysics.
Numerous experiments with ultradispersed magnetic materials discovered the dependence of such magnetic properties as coercive force, 
remanent magnetization and magnetic susceptibility on values and prehistory of the mechanical stresses applied to the samples.
The inverse problem is of independent interest: 
to determine the magnetic prehistory of the sample from its known mechanical properties. 
This task is extremely important for the magnetic measurements of fatigue of metal constructions, 
as well as in the study of the paleointensity.
Owing to the research based on the dependence of magnetic properties on mechanical stresses, 
ultradispersed magnets are used, for example, in the sensors of heavy load, in the technology of transformer cores, 
and control systems of fatigue of metal structures, 
in electronic article surveillance and many other technical developments.

\section{Model}
\begin{figure}
\centering
\includegraphics[scale=1.2]{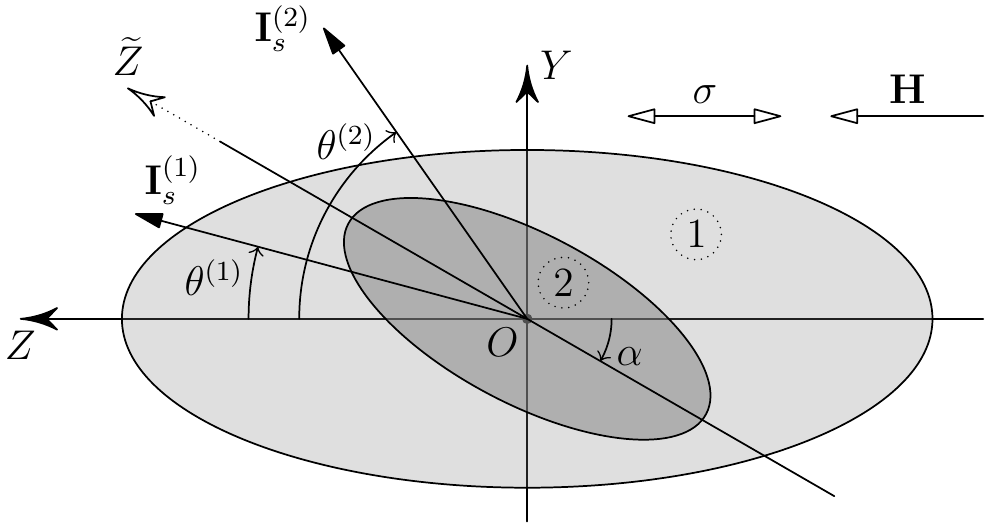}
\caption{Illustration of the model of two-phase particle}
\label{fig:1}
\end{figure}

\begin{enumerate}
\item  Homogeneously magnetized nanoparticle (phase {\em 1}) of volume $V$
has the form of ellipsoid with elongation\footnote{{\em elongation} --- 
the ratio of the length $a$ of semi-major axis of the ellipsoid to the length $b$ of semi-minor one}  $q_1$, 
and its long axis oriented along the $Oz$-axis (see fig.~\ref{fig:1}).

\item Nanoparticle contains an uniformly magnetized ellipsoidal inclusion (phase {\em 2}) 
with a volume $v=\varepsilon V$ and elongation of $q$.

\item The angle between the long axes of the particle and the inclusion is $\alpha$. 

\item It is considered that the axes of crystallographic anisotropy of both uniaxial ferromagnets are parallel 
to the long axes of the ellipsoids, and the vectors of spontaneous magnetization of 
phases ${{\mathbf I}}^{(1)}_s$ and ${{\mathbf I}}^{(2)}_s$ lie in the plane $yOz$, 
that contains the long axes of the magnetic phases, and make angles $\theta^{(1)}$ and $\theta^{(2)}$ 
with the $Oz$ axis, respectively.

\item Both external magnetic field $H$ and uniaxial mechanical stresses $\sigma $ are applied along the $Oz$-axis. 

\item The volume of nanoparticles exceeds the volume of superparamagnetic transition.
\end{enumerate}

\section{Results}
\begin{figure}[p]
\centering
\includegraphics[scale=0.96]{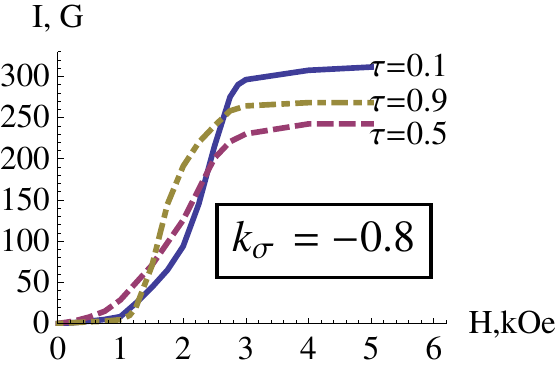}
\includegraphics[scale=0.96]{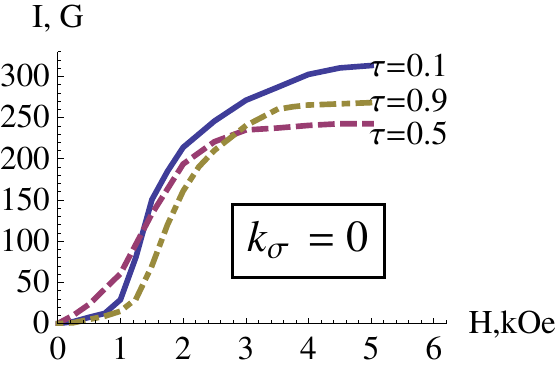}
\includegraphics[scale=0.96]{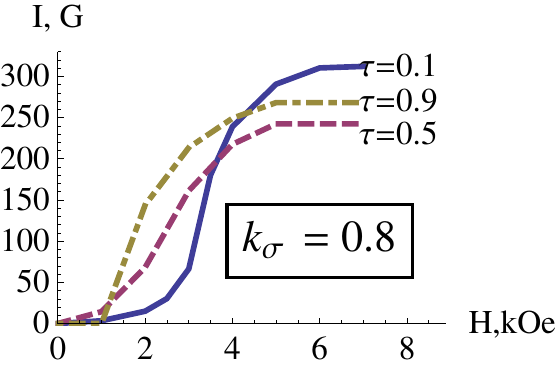}
\caption{The effect of mechanical stresses $k_{\sigma}=3\lambda_{100}\sigma/K_A$ and the relative volume of the cobalt
    coating $\tau=1-\varepsilon$ on the magnetization of elongated nanoparticles 
    (elongation $q_1=3$, constant of interfacial exchange interaction $A_{in}=0$, 
    $\lambda_{100}$ and $K_A$ are magnetostriction constant and anisotropy constant, respectively)}
\label{fig:2}
\end{figure}

The calculation of the magnetization, held within the framework of the above mentioned model 
of two-phase particles~--- maghemite ($\gamma$-$Fe_2O_3$) epitaxially coated with 
cobalt ferrite ($CoFe_2O_4$)~--- is shown in fig.~\ref{fig:2}.
It is easy to see that stretching shifts the magnetization curves to lower magnetic fields~$H$, 
and the compression leads to the opposite effect. 
At the same time mechanical stresses do not affect the saturation magnetization, 
which is determined by the thickness of cobalt coating. 
These results are determined by the dependence of the critical fields of magnetization reversal on the stresses: 
stretching decreases the critical fields of magnetization reversal and compression increases it, 
and the coercivity of the particles changes consequently.

\begin{figure}[p]
\includegraphics[scale=0.90]{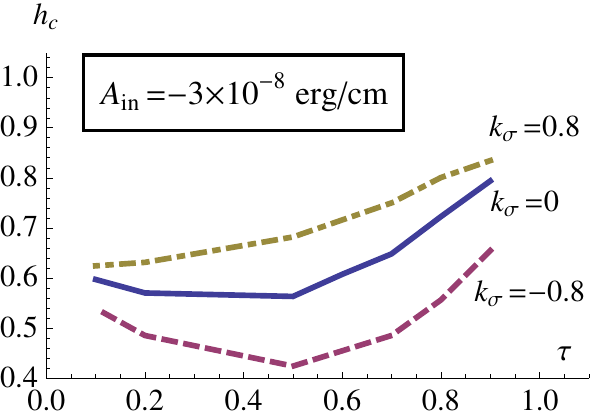}
\includegraphics[scale=0.90]{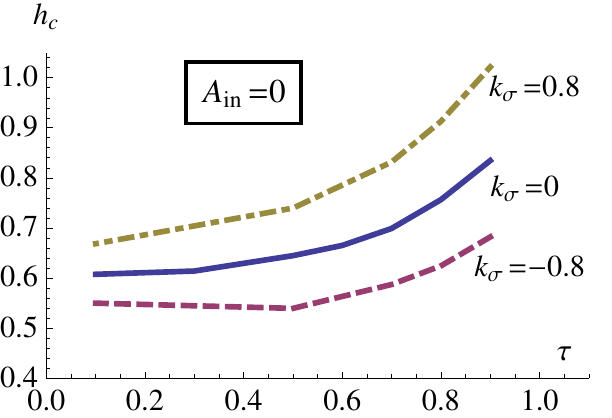}
\includegraphics[scale=0.90]{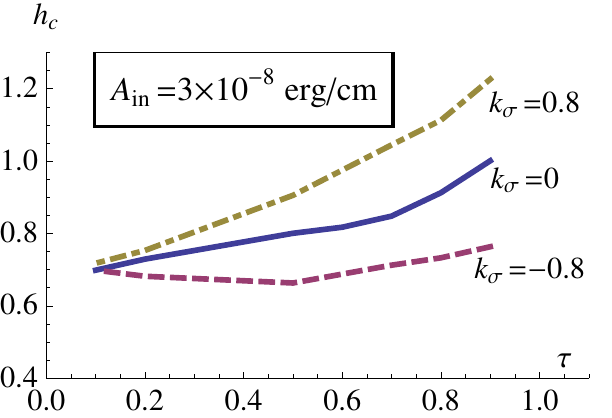}
\caption{Dependence of the relative coercive force $h_c = H_c/H_{c1}$ of elongated nanoparticles 
    on the relative volume of cobalt coating $\tau$ and on the value of the exchange interaction 
    through the interface $A_{in}$ if $q_1 = 3$, $H_{c1} = 2947$ erg}
\label{fig:3}
\end{figure}

Coercivity $H_c$ depends not only on stress but also on the magnitude of the exchange interaction 
through the interface $A_{in}$ and on the relative amount of cobalt coating $\tau=1-\varepsilon$. 
When $A_{in}=0$ or $A_{in}=3\times10^{-8}$ erg/cm coercivity of the particles increases monotonically 
with an increase in the relative volume of coating, 
whereas for $A_{in} =-3\times10^{-8}$ erg/cm behavior of the $H_c$ is nonmonotonic (see fig.~\ref{fig:3}).

\begin{figure}[p]
\includegraphics[scale=0.90]{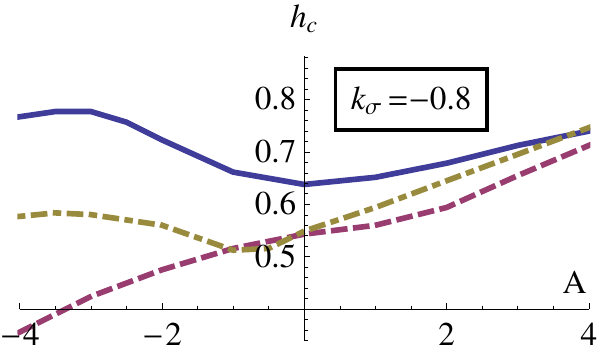}
\includegraphics[scale=0.90]{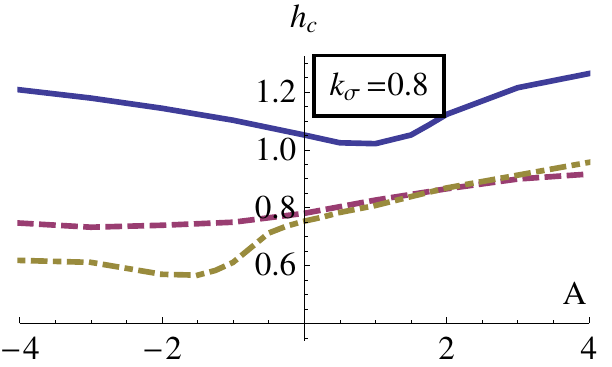}
\includegraphics[scale=0.90]{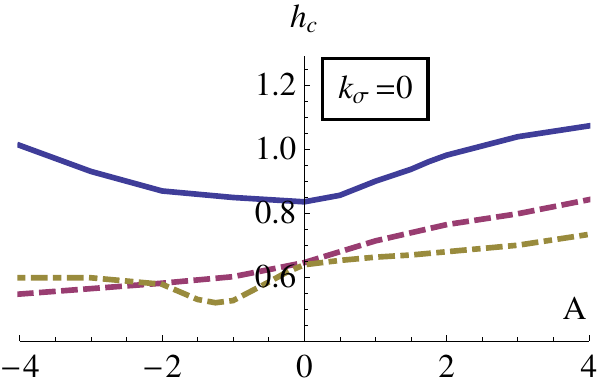}
\caption{Dependence of the relative coercive force $h_c = H_c/H_{c1}$ of elongated nanoparticles 
    on the value of the exchange interaction through the interface $A_{in}$
    and on the relative volume of cobalt coating $\tau$ (solid curve corresponds to $\tau = 0.9$, dashed
    curve --- to $\tau = 0.5$, dot-dashed curve --- $\tau = 0.1$) 
    if $q_1 = 3$, $H_{c1} = 2947$ erg}
\label{fig:4}
\end{figure}

In addition, a negative exchange interaction leads to a decrease in the coercive force~$H_c$ 
as compared with $A_{in} = 0$, and positive --- to its increasing. 
Features of the dependence of the coercive force of the interfacial exchange interaction $A_{in}$ shown in fig.~\ref{fig:4}.
One can see that the nonmonotonic behavior $H_c = H_c (A)$ is characteristic 
of nanoparticles with a large ($\tau = 0.9$) or small ($\tau = 0.1$) thickness of the cobalt coating 
and is being implemented in both positive and negative values of $A_{in}$.

The nonmonotonic behavior of the coercive force can be explained by considering the ratio of 
the exchange interaction $A_{in}$ and interfacial magnetostatic interaction $A_{ms}$.
While $A_{in}>A_{ms}$, the coercivity of the system is determined by the critical field 
of magnetization reversal from a state with a parallel orientation of the magnetic moments of the phase of nanoparticles 
to the state in which one phase (conventionally, {\em the first}) is antiparallel to the external magnetic field~$H$ 
and the other parallel to it.
According to~\cite{Geocosmos2010}, this critical field decreases with increasing of $A_{in}$.
When $A_{in}<A_{ms}$, nanoparticles with parallel magnetic moments of the phases remagnetized 
by switching of the magnetic moment of the first phase into a state parallel to the field $H$.
The critical field of magnetization reversal in this case should increase with increasing of $A_{in}$.

Noted above nonmonotonic behavior of $H_c$ did not observed in~\cite{Sawatzky1969,Yang1994}, 
which is obviously associated with a narrower, than in the present study, 
spectrum of the critical fields of magnetization reversal of two-phase particles. 
A qualitative comparison of the results with similar calculations presented in~\cite{Yang1994,Yang1991,Aharoni1988}
shows that, just as in these papers, with the growth of phase $CoFe_2O_4$, coercivity increases up to saturation.

\bibliographystyle{ieeetr}
\bibliography{xiamen_heterophase}
\end{document}